\documentclass{PoS}
\usepackage{mathptmx}
\usepackage{calrsfs}
\usepackage{pdfpages}
\DeclareMathAlphabet{\mathcal}{OMS}{zplm}{m}{n}

\let\OLDthebibliography\thebibliography
\renewcommand\thebibliography[1]{
  \OLDthebibliography{#1}
  \setlength{\parskip}{0pt}
  \setlength{\itemsep}{-3pt}
\footnotesize
}

\title{Studying parton correlations via double parton scatterings in associated quarkonium production at the LHC and the Tevatron}

\ShortTitle{Parton correlations via DPS in associated quarkonium production}

\author{\speaker{Nodoka~Yamanaka}\thanks{Supported by JSPS Postdoctoral Fellowships for Research Abroad.}\\
        IPNO, CNRS-IN2P3, Univ. Paris-Sud, Universit\'e Paris-Saclay, 
91406 Orsay Cedex, France\\
        E-mail: \email{yamanaka@ipno.in2p3.fr}
}

\author{Jean-Philippe Lansberg\\
        IPNO, CNRS-IN2P3, Univ. Paris-Sud, Universit\'e Paris-Saclay, 
91406 Orsay Cedex, France\\
}

\author{Hua-Sheng Shao\\
Laboratoire de Physique Th\'eorique et Hautes Energies (LPTHE), UMR 7589, Sorbonne Universit\'e et CNRS, 4 place Jussieu, 75252 Paris Cedex 05, France\\
}

\author{Yu-Jie Zhang\\
        Beijing Key Laboratory of Advanced Nuclear Energy Materials and Physics, and School of Physics, Beihang University, Beijing 100191, China\\
}

\abstract{
Quarkonium production in proton-proton ($pp$) collision provides interesting means to study the parton content and their correlations in the proton. 
Recent LHC and Tevatron data of $J/\psi + Z$, $J/\psi + W$ and $J/\psi +J/\psi$ production suggest the relevance of double parton scatterings (DPSs) as opposed to single parton scatterings (SPSs). 
In this proceedings contribution, we review the corresponding SPS contributions and discuss their upper limits set up by quark-hadron duality. 
These allow us to perform an improved extraction of the DPS yields and of their implications.
}

\FullConference{The 39th International Conference on High Energy Physics (ICHEP2018)\\
		4-11 July, 2018\\
		Seoul, Korea}

\begin{document}


\section{Introduction}

\vspace{-1em}

The quarkonium production at colliders is studied to probe the perturbative and nonperturbative properties of QCD.
To test the nonrelativistic QCD (NRQCD) by looking for the color-octet (CO) contribution, the production of $J/\psi +W$ may be used as a golden channel \cite{Barger:1995vx}.
The final state $\Upsilon +W$ could be a decay channel of the charged Higgs boson \cite{Grifols:1987iq} which is a signature of physics beyond the standard model.
The quarkonium + $\gamma$ production was proposed to constrain the quarkonium-production~\cite{Roy:1994vb,Mathews:1999ye,Li:2008ym,Lansberg:2009db,Li:2014ava} and to study the gluon distribution in the proton~\cite{Doncheski:1993dj,Dunnen:2014eta}.
The di-$J/\psi$ production is important in the study of double parton scatterings (DPS) \cite{Kom:2011bd,Baranov:2015cle,Borschensky:2016nkv,Lansberg:2014swa} and to extract the distribution of linearly polarized gluons in the proton~\cite{Lansberg:2017dzg}.
Quantifying the DPS is also important in the search for new particles and interactions, since it may become a background in multi-particle final states of $pp$ collisions at high-energy.
On the experimental side, new data of quarkonium associated productions recently became available.
The $J/\psi +W$ \cite{Aad:2014rua} and $J/\psi +Z$ \cite{Aad:2014kba} final states were observed by the ATLAS Collaboration.
The D0 \cite{Abazov:2014qba}, CMS \cite{Khachatryan:2014iia}, ATLAS \cite{Aaboud:2016fzt}, and LHCb \cite{Aaij:2011yc,Aaij:2016bqq} Collaborations also studied the $J/\psi +J/\psi$ production, while D0 \cite{Abazov:2015fbl} and CMS \cite{Khachatryan:2016ydm} analyzed $J/\psi +\Upsilon$ and $\Upsilon + \Upsilon$, respectively.
On the other hand, the single-parton scattering (SPS) contributions for $J/\psi+W$, $J/\psi+Z$, $J/\psi+J/\psi$, and $J/\psi+\Upsilon$ were theoretically computed in NRQCD \cite{Li:2010hc,Mao:2011kf,Gong:2012ah,Lansberg:2013wva,Lansberg:2016muq,Shao:2016wor}.
In this proceedings contribution, we present the result of the SPS calculation of $J/\psi +W$, $J/\psi +Z$, and $J/\psi +J/\psi$ \cite{Lansberg:2014swa,Lansberg:2013qka,He:2015qya,Sun:2014gca} production in the color evaporation model (CEM).

\vspace{-1em}

\section{Analysis of the ATLAS data for $J/\psi +Z$ and $J/\psi +W$ productions in the CEM}

\vspace{-1em}

\begin{figure}[hbt!]
\vspace{-1em}
\begin{center}
\includegraphics[width=0.5\columnwidth]{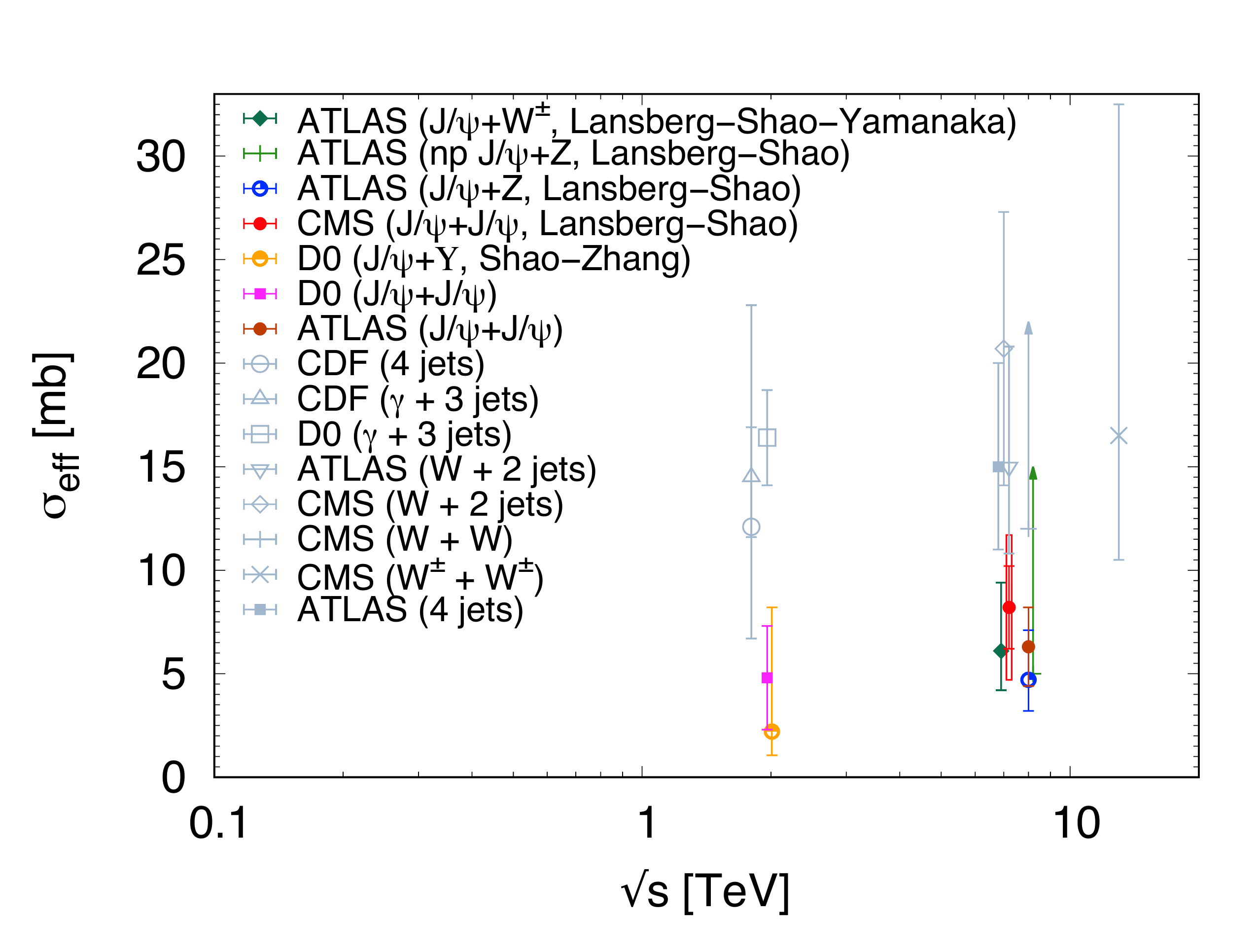}
\vspace{-1em}
\caption{
Summary of several extractions of $\sigma_{\rm eff}$. 
Quarkonium related extractions are shown in color.
}
\label{fig:sigmaeff_summary}
\end{center}
\vspace{-2em}
\end{figure}

Although multiple parton interactions are of higher twist, they become important at high energies.
It is expected to restore the unitarity of the cross section, enhanced by the parton flux which increases with the energy, 
Let us discuss here the DPS which is the simplest case.
By assuming that the two partons are uncorrelated, the DPS cross section is often parametrized  with the formula $\sigma_{\rm DPS} (A+B) = \frac{1}{1+\delta_{AB}} \frac{\sigma (A) \sigma (B)}{\sigma_{\rm eff}}$, where $\delta_{AB} =1$ corresponds to the case where we have $A=B$ in the final state.
In Fig. \ref{fig:sigmaeff_summary}, we summarize the current situation of $\sigma_{\rm eff}$ from different extractions.

To look for signs of DPS, one can estimate the upper limit of the SPS contribution and subtract it from the experimental data of ATLAS, so as to establish the residual part which should be the DPS.
An adequate approach is to use the CEM, based on quark-hadron duality, where the quarkonium is considered as a state of quark-antiquark pair with its invariant mass below the open-heavy flavor threshold.
The CEM cross section is given as $\sigma^{\rm (N)LO,\ \frac{direct}{prompt}}_{ J/\psi} = {\cal P}^{\rm (N)LO,\frac{direct}{prompt}}_{J/\psi}\int_{2m_c}^{2m_D} \frac{d\sigma_{c\bar c}^{\rm (N)LO}}{d m_{c\bar c}}d m_{c\bar c},$ where ${\cal P}^{\rm (N)LO,{prompt}}_{J/\psi} = 0.014$ (LO), 0.009 (NLO) \cite{Lansberg:2016rcx} is supposed to be a universal parameter. 
In the CEM, the single-quarkonium production overshoots the experimental data at high transverse momentum $p_T$ \cite{Lansberg:2006dh,Andronic:2015wma,Lansberg:2016rcx}, owing to the dominance of the gluon fragmentation topologies.
We expect this to also occur for $J/\psi +W$ and $J/\psi +Z$ productions.
We can therefore consider the CEM calculations as a conservative upper limit on the SPS cross section.
It can be calculated at NLO using {\small \sc MadGraph5\_aMC@NLO} \cite{Alwall:2014hca}.

\begin{table}[htb]
\vspace{-1em}
\begin{center}
\caption{
Comparison of the NLO CEM result with the 
experimental data of ATLAS.
The error bars are the combined statistical and systematic uncertainties.
}
\begin{tabular}{l|ll}
\hline
& ATLAS & CEM \\
\hline
$J/\psi +Z$ & $1.6 \pm 0.4$ pb \cite{Aad:2014kba} & $0.19^{+0.05}_{-0.04}$ pb \cite{Lansberg:2016rcx}\\
$J/\psi +W$ & $4.5 ^{+1.9}_{-1.5}$ pb \cite{Aad:2014rua} & $0.28 \pm 0.07$ pb \cite{Lansberg:2017chq}\\
\hline
\end{tabular}
\label{table:comparison}
\vspace{-2em}
\end{center}
\end{table}

\begin{figure}[htb]
\vspace{-1em}
\begin{center}
\includegraphics[width=.35\columnwidth]{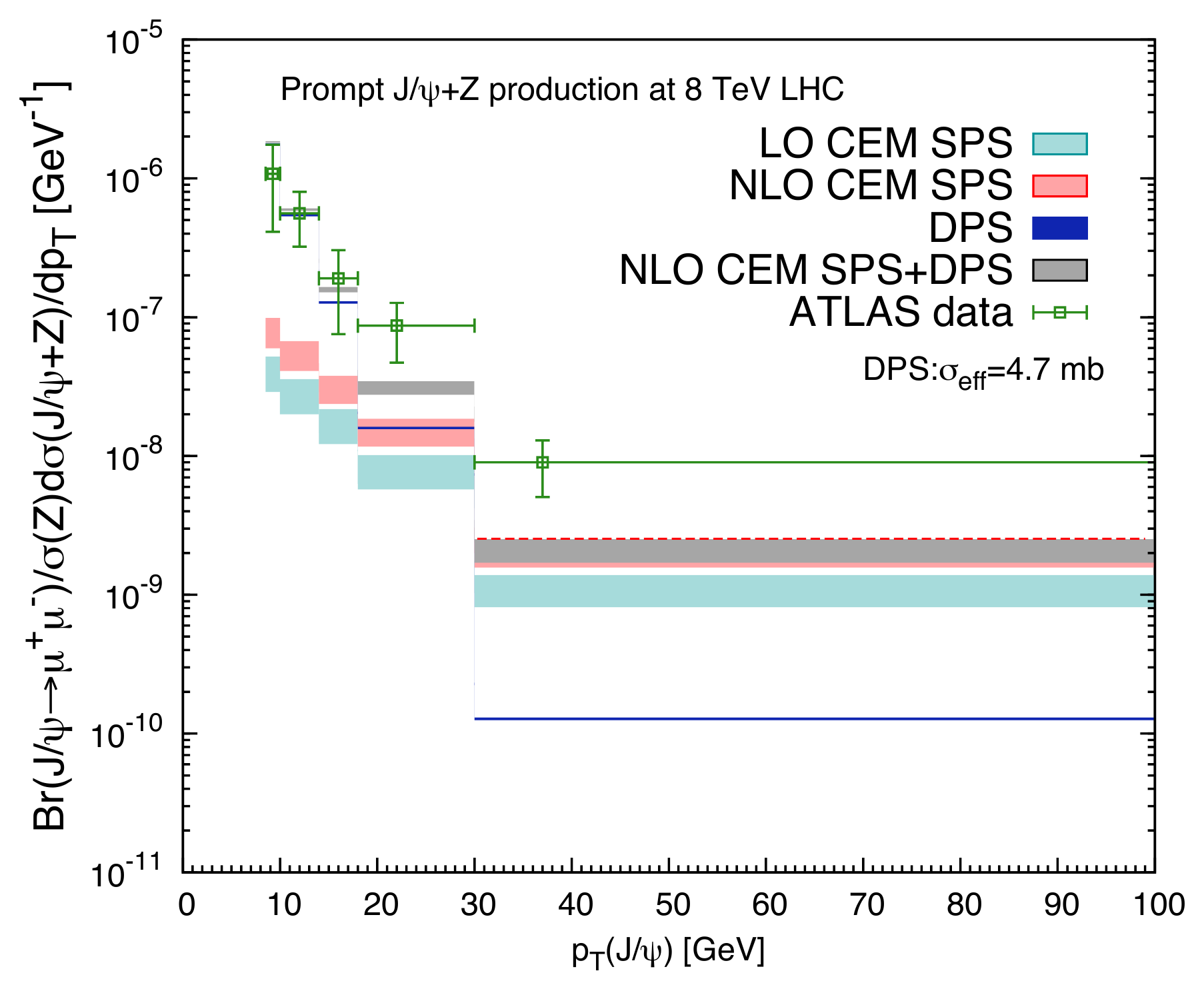}
\includegraphics[width=.35\columnwidth]{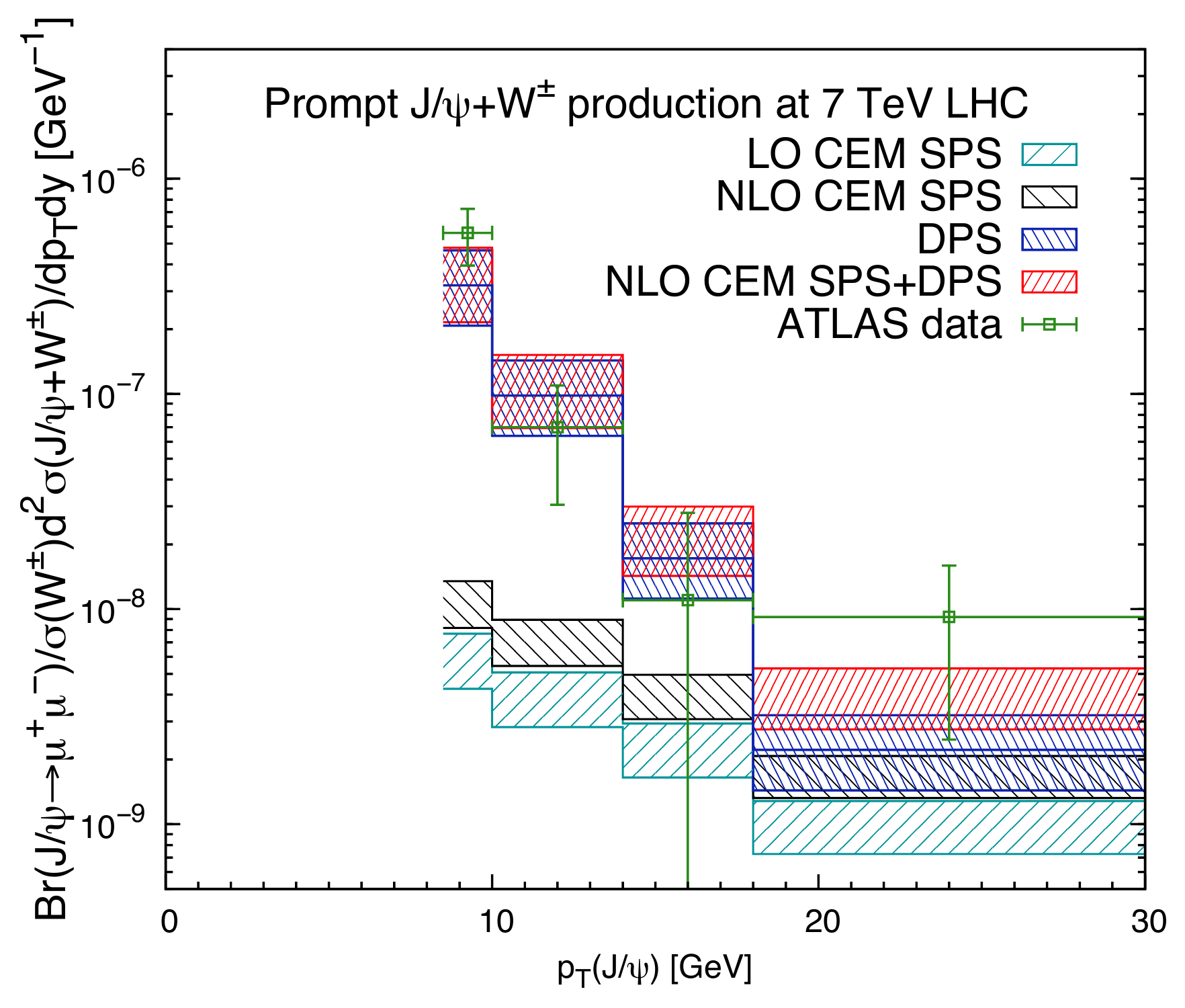}
\vspace{-1em}
\caption{
The $p_T$ distribution of the $J/\psi$ in the $J/\psi +Z$ \cite{Lansberg:2016rcx} and $J/\psi +W$ \cite{Lansberg:2017chq} production cross section in the CEM.
We also plot the ATLAS experimental data \cite{Aad:2014kba,Aad:2014rua} for comparison.
}
\label{fig:pT_distribution}
\end{center}
\vspace{-2em}
\end{figure}

We show the results for the $J/\psi +Z$ and $J/\psi +W$ productions in Table \ref{table:comparison}.
We see that the upper limits given by the CEM is significantly smaller than the ATLAS experimental data.
This difference is particularly accentuated in the low $p_T$ region (see Fig. \ref{fig:pT_distribution}).
These results show that the SPS alone is not sufficient to explain the ATLAS data.

By assuming that the DPS fills this gap, we fit $\sigma_{\rm eff}$ to the ATLAS total inclusive cross section measurements the SPS contribution, where the $p_T$ distribution is shown in Fig. \ref{fig:pT_distribution}.
From this fit, we obtain $\sigma_{\rm eff} = (4.7 ^{+2.4}_{-1.5})$ mb \cite{Lansberg:2016rcx} for the ${J/\psi +Z}$ and $\sigma_{\rm eff} = (6.1 ^{+3.3}_{-1.9})$ mb \cite{Lansberg:2017chq} for the ${J/\psi +W}$ productions.

\vspace{-1em}

\section{$J/\psi + J/\psi $ in the CEM}

\vspace{-1em}

\begin{figure}[htb]
\vspace{-1em}
\begin{center}
\includegraphics[width=.35\columnwidth]{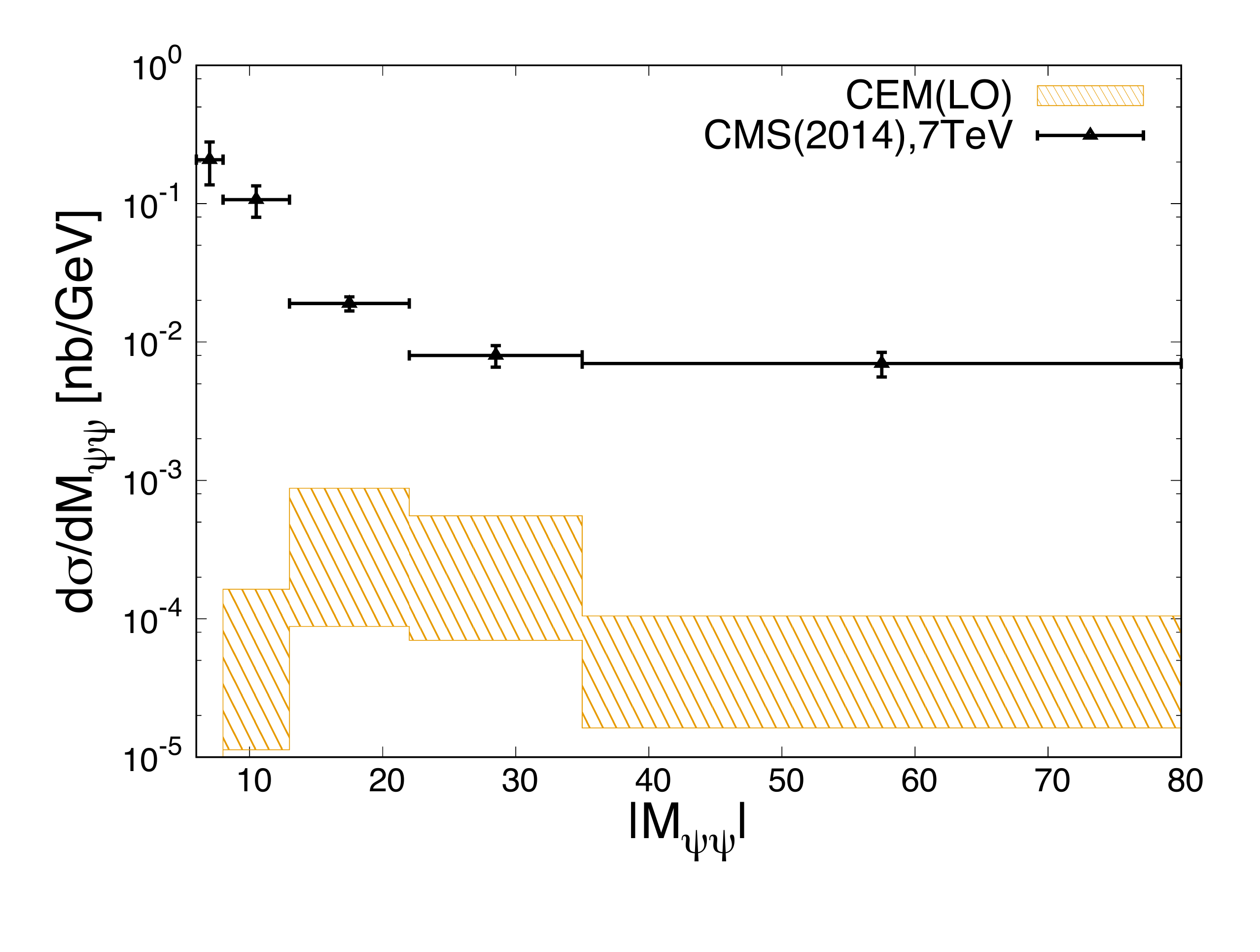}
\includegraphics[width=.35\columnwidth]{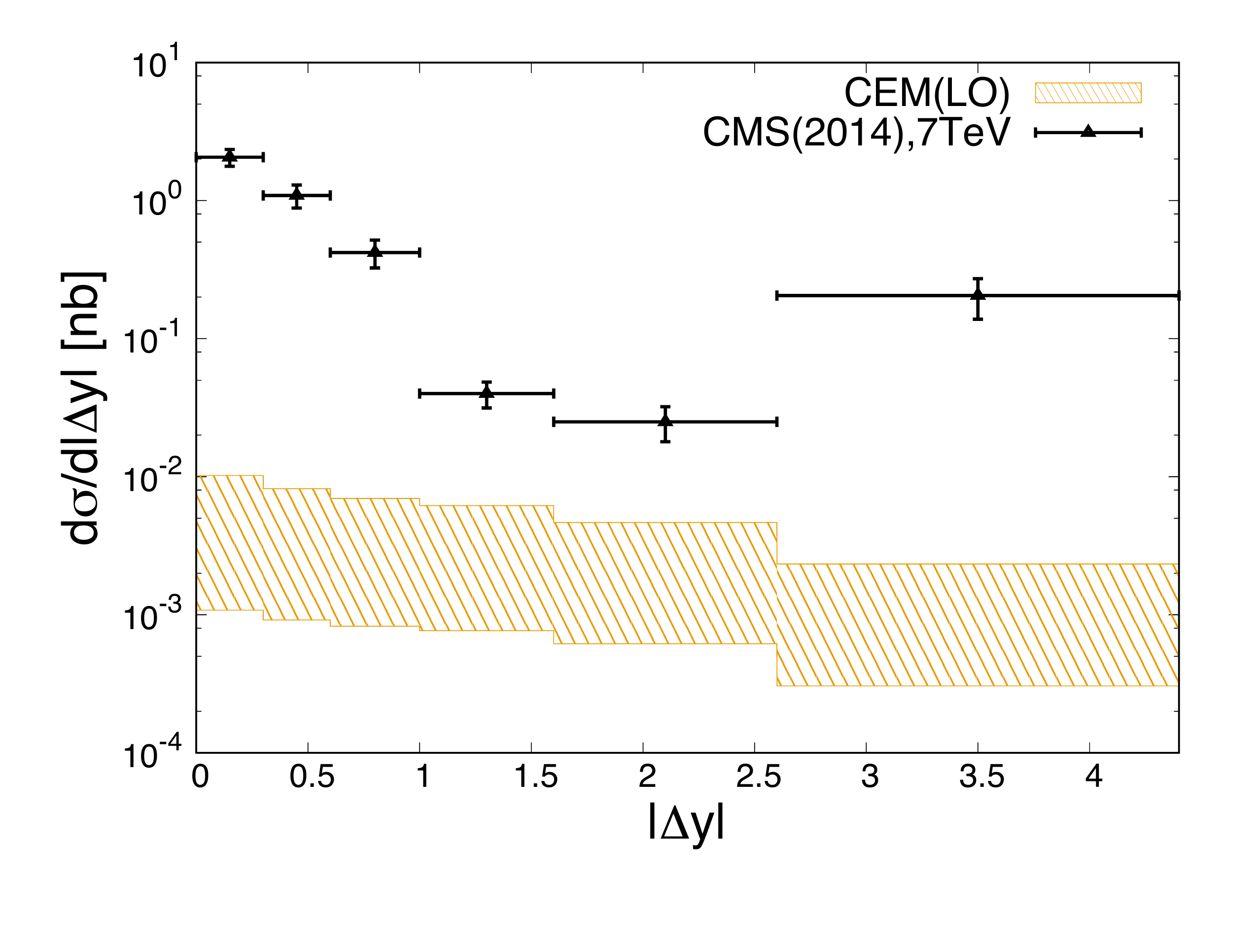}
\vspace{-2em}
\caption{
The invariant mass (left panel) and $\Delta y$ (right panel) distributions of di-$J/\psi$ production (CMS setup, $\sqrt{s} = 7$ TeV).
}
\label{fig:double_Jpsi_CMS7TeV}
\end{center}
\vspace{-3em}
\end{figure}

Since the SPS amplitude in the CEM is dominated by the color-octet contribution at large transverse momentum region, it might also be used to give an estimation of the CO contributions.
This is particularly interesting to investigate the large invariant mass $M_{\psi \psi}$ and $\Delta y$ regions of the di-$J/\psi$ production where the experimental data of CMS is showing an enhancement of the cross section \cite{Khachatryan:2014iia}.
In Fig. \ref{fig:double_Jpsi_CMS7TeV}, we show our result of the LO CEM analysis.
We see no particular enhancement from the CEM at large $M_{\psi \psi}$ and $\Delta y$.
Our result is thus supporting the dominance of the DPS in di-$J/\psi$ production in these regions, extracted from the CMS \cite{Khachatryan:2014iia} ($\sigma_{\rm eff} = (8.2 \pm 2.0_{\rm stat} \pm 2.9_{\rm sys})$ mb \cite{Lansberg:2014swa}), D0 ($\sigma_{\rm eff} = (4.8 \pm 0.5_{\rm stat} \pm 2.5_{\rm sys})$ mb) \cite{Abazov:2014qba}, and ATLAS Collaborations ($\sigma_{\rm eff} = (6.3 \pm 1.6_{\rm stat} \pm 1.0_{\rm sys})$ mb) \cite{Aaboud:2016fzt}.

\vspace{-1em}

\section{Conclusion}

\vspace{-1em}

To conclude, we analyzed in the CEM the production processes of $J/\psi + W/Z$ (NLO) and $J/\psi + J/\psi$ (LO). 
In this approach, it is possible to set an upper limit on the SPS contribution and to extract the DPS from the experimental data.
From the ATLAS data, we could see the evidence of the DPS with $\sigma_{\rm eff} = (4.7 ^{+2.4}_{-1.5})$ mb (${J/\psi +Z}$), and $\sigma_{\rm eff} = (6.1 ^{+3.3}_{-1.9})$ mb (${J/\psi +W}$), which are compatible with $\sigma_{\rm eff}$ extracted from the di-$J/\psi$ production and other central rapidity quarkonium data.
The enhancement at large $\Delta y$ and invariant mass of the $J/\psi +J/\psi$ production is explained by $\sigma_{\rm eff}<10$ mb.

\vspace{-1em}



\bibliographystyle{Science}

\bibliography{yamanaka}

\end{document}